\documentclass[twocolumn,showpacs,aps,prl,superscriptaddress,floatfix]{revtex4}

\usepackage{epsfig}
\usepackage{graphics}
\usepackage{graphicx}
\usepackage{epic}
\usepackage{floatflt}
\usepackage{dcolumn}
\usepackage{amsmath}

\long\def\inst#1{\par\nobreak\kern 4pt\nobreak
    {\it #1}\par\vskip 10pt plus 3pt minus 3pt}

\def\sss{\scriptscriptstyle}
\def\barpd{{\raise.35ex\hbox{${\sss (}$}}--{\raise.35ex\hbox{${\sss )}$}}}
\def\dbarp{\hbox{$D^{0}$\kern-1.3em\raise1.5ex\hbox{\barpd}}}
\def\dbarpnozero{\hbox{$D$\kern-0.85em\raise1.5ex\hbox{\barpd}}}

\newcommand{\BaBarYear}      {05}
\newcommand{\BaBarNumber}    {26}
\newcommand{\SLACPubNumber}  {11324}

\input pubboard/babarsym

\def\Dcp    {\ensuremath{D^{0}_{\CP }}\xspace}
\def\Acp    {\ensuremath{{\cal A}_{\CP }}\xspace}
\def\Acpp   {\ensuremath{{\cal A}_{\CP+ }}\xspace}
\def\Acpm   {\ensuremath{{\cal A}_{\CP- }}\xspace}
\def\Acppm  {\ensuremath{{\cal A}_{\CP \pm }}\xspace}
\def\Rcp    {\ensuremath{{\cal R}_{\CP }}\xspace}
\def\Rcpp   {\ensuremath{{\cal R}_{\CP+ }}\xspace}
\def\Rcpm   {\ensuremath{{\cal R}_{\CP- }}\xspace}
\def\Rcppm  {\ensuremath{{\cal R}_{\CP \pm }}\xspace}
\def\cpp    {\ensuremath{\CP+ }\xspace}
\def\cpm    {\ensuremath{\CP- }\xspace}

\def\de {\ensuremath{{\rm \Delta}E}\xspace}

\begin{document}

\begin{flushleft}
\babar-PUB-\BaBarYear/\BaBarNumber \\
SLAC-PUB-\SLACPubNumber \\
\end{flushleft}

\title{Measurement of \boldmath{\CP}~Observables for the Decays \boldmath{$\Bpm\to\Dcp\Kstarpm$}}

%
\author{B.~Aubert}
\author{R.~Barate}
\author{D.~Boutigny}
\author{F.~Couderc}
\author{Y.~Karyotakis}
\author{J.~P.~Lees}
\author{V.~Poireau}
\author{V.~Tisserand}
\author{A.~Zghiche}
\affiliation{Laboratoire de Physique des Particules, F-74941 Annecy-le-Vieux, France }
\author{E.~Grauges}
\affiliation{IFAE, Universitat Autonoma de Barcelona, E-08193 Bellaterra, Barcelona, Spain }
% ...
\collaboration{The \babar\ Collaboration}
\noaffiliation

\vspace{5mm}

\date{\today}

\begin{abstract}
\noindent
Using a sample of 232 million $\FourS\to\BB$ events collected with the
\babar\ detector at the \pep2 \BF\ in 1999--2004, 
we study $\Bm\to\Dz\Kstar$(892)$^-$ decays where $\Kstarm\to\KS\pim$ and
$\Dz\to\Km\pip$, $\Km\pip\piz$, $\Km\pip\pip\pim$ (non-\CP final states); 
\KpKm, $\pip\pim$ (\cpp eigenstates); $\KS\piz$, $\KS\phi$ and $\KS\omega$ (\cpm eigenstates). 
We measure four observables that are sensitive to the angle $\gamma$ of the CKM unitarity triangle;
the partial-rate charge asymmetries \Acppm and
the ratios of the \B-decay branching fraction in $\CP\pm$ and non-\CP decays \Rcppm: 

\begin{eqnarray}
\vspace*{-0.2in}
{\cal A}_{\CP+} &=& -0.08 \pm 0.19 \ {\rm (stat.)} \pm 0.08 \ {\rm (syst.)} \nonumber\\
{\cal A}_{\CP-} &=& -0.26 \pm 0.40 \ {\rm (stat.)} \pm 0.12 \ {\rm (syst.)} \nonumber\\
{\cal R}_{\CP+} &=& ~~1.96\pm 0.40 \ {\rm (stat.)} \pm 0.11 \ {\rm (syst.)} \nonumber\\
{\cal R}_{\CP-} &=& ~~0.65\pm 0.26 \ {\rm (stat.)} \pm 0.08 \ {\rm (syst.)} \nonumber
\end{eqnarray}

\end{abstract}

\pacs{13.25.Hw, 14.40.Nd}

\maketitle

The measurement of \CP violation in \B-meson decays offers a means to over-constrain the
unitarity triangle. A theoretically clean determination of the angle 
$\gamma$=arg($-V_{ud}V^*_{ub}/V_{cd}V^*_{cb}$) is provided by
the $\Bm\ra D^{(*)0}K^{(*)-}$ decay channels in which the 
favored $b \ra c \ubar s$ and suppressed $b \ra u \cbar s$ tree
amplitudes interfere~\cite{bib:theory,bib:gronau}.
Results on the $\Bm \to D^{(*)0}K^-$ decays have been published by the 
\babar~\cite{Aubert:2003uy,Aubert:2004fb,Aubert:2004hu} and 
BELLE~\cite{Swain:2003yu,Poluektov:2004mf} collaborations.
In this paper, we present a study based on
%
the interference between $\Bm\ra\Dz \Kstar\mathrm{ (892)}^-$ and $\Bm\ra\Dzb\Kstarm$ when both \Dz and \Dzb decay
to the same \CP\ eigenstate (\Dcp). Reference to a charge conjugate mode is implied throughout this paper unless otherwise stated. \par
We follow~\cite{bib:gronau,Gronau:2002mu} and define:

\begin{eqnarray}
\Rcppm &=& 2 \frac{\Gamma(\Bm\to D^{0}_{\CP\pm} \Kstarm) + \Gamma(\Bp\to D^{0}_{\CP\pm} \Kstarp)}
                                               {\Gamma(\Bm\to\Dz\Kstarm)+\Gamma(\Bp\to\Dz\Kstarp) } \nonumber\\
\Acppm &=& \frac{\Gamma(\Bm\to D^{0}_{\CP\pm} \Kstarm) - \Gamma(\Bp\to D^{0}_{\CP\pm} \Kstarp)}
                        {\Gamma(\Bm\to D^{0}_{\CP\pm} \Kstarm) + \Gamma(\Bp\to D^{0}_{\CP\pm} \Kstarp)}. \nonumber 
\end{eqnarray}
\noindent
Both \Acp\ and \Rcp\ carry \CP-violating information. Neglecting \Dz -\Dzb\ mixing, 
they can be expressed as follows:
\begin{eqnarray}
\label{eq:cpinfo-rcp}
\Rcppm	                &=& {1 \pm 2 r_{\B} \cos \delta \cos \gamma + r_{\B}^{2}}, \\
\Acppm			&=& \frac{\pm \ 2 r_{\B} \sin \delta \sin \gamma} {\Rcppm}, 
\end{eqnarray}
\noindent
where $\delta$ is the \CP-conserving strong phase difference between the $\Bm\ra\Dzb\Kstarm$ (suppressed) 
and $\Bm\ra\Dz\Kstarm$ (favored) amplitudes,  $r_{\B} \simeq 0.1$-$0.3$~\cite{Gronau:2002mu} is the magnitude of their ratio 
and $\gamma$ is the \CP-violating weak phase difference. A value close to $60^{\circ}$ is favored for $\gamma$ when one combines
all measurements related to the unitarity triangle~\cite{Charles:2004jd}. It is useful to introduce also new variables, 
\begin{eqnarray}
\label{xvariables}
x^\pm \equiv r_B\,\cos(\delta\pm\gamma),
\end{eqnarray}
which are better behaved (more Gaussian) in the region where $r_{\B}$ is small. 


To search for $\Bm\to\Dcp\Kstarm$ decays we use data collected 
with the \babar\ detector~\cite{bib:babar}
at the \pep2\ storage ring. The sample corresponds to an integrated luminosity of 211 \invfb
at the \FourS resonance (232 million \BB\ pairs) and 20.4 \invfb\ at an energy 40~\mev below the peak.\par

To reconstruct $\Bm\to\Dz\Kstarm$ decays, we select  \Kstarm\ candidates in
the $\Kstarm\to\KS\pim$, $\KS\to\pip\pim$  mode and \Dz candidates in eight decay channels,
$\Dz\to\Km\pip$, $\Km\pip\piz$, $\Km\pip\pip\pim$ (non-\CP final states); 
\KpKm, $\pip\pim$ (\cpp eigenstates); and $\KS\piz$, $\KS\phi$, $\KS\omega$ (\cpm eigenstates).
We optimize our event selection to minimize the statistical error on the signal yield, 
determined for each channel using simulated signal and background events. Particle identification 
is required for all charged particles except for the pions from \KS decays.\par
\KS candidates are formed from oppositely charged tracks assumed to be pions with a reconstructed invariant mass
within 13~\mevcc\ (four standard deviations) from the known \KS mass~\cite{bib:pdg2004}, $m_{\KS}$. 
All \KS candidates are refitted so that their invariant mass equals $m_{\KS}$ (mass constraint). For those 
retained to build a  $D^{0}_{\CP-}$ candidate the tracks are also constrained to emerge from a single vertex (vertex constraint). 
For those retained to build a \Kstarm we further require their flight direction
and length be consistent with a \KS\ coming from the interaction point.
The \KS\ candidate flight path and momentum must make an acute angle and the flight length in the
plane transverse to the beam direction must exceed its uncertainty by three standard deviations. 
\Kstarm candidates are formed from a \KS and a charged particle with a vertex constraint.
We select \Kstarm candidates which have an invariant mass within 75~\mevcc of the known value~\cite{bib:pdg2004}.
Finally, since the \Kstarm in \Bm\to\Dz\Kstarm is polarized, we require $|$cos~$ \theta_{H}| \geq 0.35$, where
$\theta_{H}$ is the angle in the \Kstarm\ rest frame between the daughter pion and the parent \B momentum. 
The helicity distribution discriminates well between a \B-meson decay and an event from the \epem \to \qqbar\ ($q\in \{u,d,s,c\}$) 
continuum, since the former is distributed as $\cos^2\theta_{H}$ and the latter is flat.\par
Some decay modes of the \Dz contain a \piz. We combine pairs of photons to form a \piz candidate with a total energy
greater than 200~\mev and an invariant mass between 115 and 150~\mevcc. A mass-constrained fit is applied to the selected
\piz\ candidates. 
Composite particles included in the \cpm\  modes are vertex constrained. 
Candidate $\phi$ ($\omega$) mesons are constructed from $\Kp\Km$ ($\pip\pim\piz$) particle combinations 
with the invariant mass required to be within  12 (20)~\mevcc\ or two standard deviations of the known values~\cite{bib:pdg2004}. 
Two further requirements are made on the $\omega$ candidates. The magnitude of the cosine of the
helicity angle between the \Dz\ momentum in the rest frame of the $\omega$ and the normal to the plane containing all three
decay pions, must be greater than 0.25. The Dalitz angle~\cite{bib:dalitzomega}, defined as the angle
between the momentum of one daughter pion in the $\omega$ rest frame and the direction of one of the other two pions in the
rest frame of the two pions, must have a cosine with a magnitude less than~0.9. \par
Except for the $\KS\piz$ final state, all \Dz candidates are mass and vertex constrained. We select \Dz\ candidates with an
invariant mass differing from the known mass~\cite{bib:pdg2004} by less than 12~\mevcc for all channels except $\KS\piz$ (30~\mevcc) and $\KS\omega$
(20~\mevcc). These limits are about twice the corresponding RMS mass resolutions. \par
To suppress the background due to \epem \to \qqbar reactions, we require $|$cos~$ \theta_B| \leq 0.9$, where
$\theta_B$ is defined as the angle between the \B candidate momentum in the \FourS\ rest frame and the beam axis.  
In \qqbar background events the $\cos\theta_B$ distribution is uniform, while for \B-mesons it follows a $\sin^2\theta_B$ distribution.
 We also use global event shape variables
to distinguish between \qqbar\ continuum events, which have a two-jet-like topology in the \FourS\ rest frame, and \BB\ events, which
are more spherical. We require $|$cos~$ \theta_{T}| \leq 0.9$ where $\theta_{T}$ is
the angle between the thrust axes of the \B candidate and that of the rest of the event.
We construct a linear discriminant~\cite{bib:fisher} from cos~$\theta_{T}$ and Legendre monomials~\cite{bib:muriel}
describing the energy flow in the rest of the event. \par
We identify \B candidates using two nearly independent kinematic variables: the beam-energy-substituted mass
$\mes=\sqrt{(s/2+{\bf p_0 \cdot p_B})^2/E_0^2-p_B^2}$ and the energy difference $\Delta E=E_B^*-\sqrt{s}/2$, where $E$ and $p$ are energy and momentum, the
subscripts 0 and $B$ refer to the \epem-beam system and the \B candidate respectively; $s$ is the square of the
center-of-mass (CM) energy and the asterisk labels the CM frame. For signal modes, the \mes\ distributions are all described by the same
Gaussian function $\mathcal{G}$ centered at the \B mass with a 2.6~\mevcc\ resolution (for \Dz \to \KS\piz the peak is slightly wider, 2.7~\mevcc).
The \de distributions are centered on zero for signal with a resolution of 11 to 13~\mev\ for all channels except 
\KS\piz\ for which the resolution is asymmetric and about 30~\mev. We define a signal region through the requirement 
$|\Delta E| < 50(25)$~\mev\ for \KS\piz (all other modes).\par

A background for $\Bm\to\Dz(\pip\pim)\Kstarm(\KS\pim)$ is the decay $\Bm\to\Dz(\KS \pip\pim)\pim$ which contains 
the same final state as the signal but has a branching fraction 600 times larger. 
We therefore explicitly veto any selected \B~candidate containing a $\KS\pip\pim$ combination within 25~\mevcc~of the \Dz~mass. No background remains.\par

In those events where we find more than one acceptable \B candidate 
(less than $25$\% of selected events depending on the \Dz mode), 
we choose that with the smallest $\chi^2$ formed from the differences of the measured and true \Dz and \Kstarm\
masses scaled by the mass spread which includes the resolution and, for the \Kstarm, the natural width. 
Simulations show that no bias is introduced by this choice and the correct candidate is picked at least 82\% of the time. \par

According to simulation of signal events, the total reconstruction efficiencies are:
13.1\% and 14.2\% for the \cpp modes $\Dz\to\KpKm$  and $\pip\pim$;
5.5\%, 10.0\% and 2.4\%  for the \cpm modes $\Dz\to\KS\piz$, $\KS\phi$ and $\KS\omega$;
13.3\%, 4.3\% and 8.2\% for the  non-\CP modes $\Dz\to\Km\pip$, $\Km\pip\piz$ and $\Km\pip\pip\pim$. \par

To study \BB\ backgrounds we look in sideband regions away from the signal region in \de and  $m_{\Dz}$. 
We define a \de\ sideband
in the  interval $-100 \leq \de \leq -60 $ and $60 \leq \de \leq 200 \mev $ for all modes
except $\Dz \to \KS \piz$ for which the inside limit is $\pm95$ rather than 60~\mev. 
The sideband region in $m_{\Dz}$ is defined by requiring that this quantity differs from the \Dz mass peak by more than four standard deviations.
It provides sensitivity to doubly-peaking background sources which mimic signal both in \de\ and \mes. This pollution comes from
either charmed or charmless \B-meson decays that do not contain a true \Dz. As many of the possible contributions
to this background are not well known, we attempt to measure its size by including the $m_{\Dz}$ sideband in the fit described below.\par

An unbinned extended maximum likelihood fit to \mes distributions in the range $5.2\leq\mes\leq 5.3$~\gevcc 
is used to determine yields and \CP-violating quantities \Acp\ and \Rcp.
We use the same Gaussian function $\mathcal{G}$ to describe the signal shape for all modes considered. The 
combinatorial background in the \mes\ distribution is modeled with a threshold function~\cite{bib:argus} $\mathcal{A}$.
Its shape is governed by one parameter $\xi$ that is
left free in the fit.
We fit simultaneously \mes distributions of nine samples: the non-\CP , \cpp\
and \cpm samples for ({\it i}) the signal region, ({\it ii}) the $m_{\Dz}$ sideband and ({\it iii}) the \DeltaE sideband.
We fit three probability density functions (PDF) weighted by the unknown event yields. 
For the \DeltaE sideband, we use $\mathcal{A}$. For the $m_{\Dz}$ sideband~(sb) we use $ a_{{\rm sb}} \cdot
\mathcal{A}$ + $b_{{\rm sb}} \cdot \mathcal{G}$, where $\mathcal{G}$ accounts for the doubly-peaking \B decays. 
For the signal region PDF , we use $a \cdot \mathcal{A}+ b \cdot \mathcal{G}+ c \cdot \mathcal{G}$,
where $b=N_{{\rm peak}}$ is scaled from $b_{{\rm sb}}$ according to the ratio of the $m_{\Dz}$ signal-window to sideband
widths and $c$ is the number of $\Bpm\to\Dz\Kstarpm$ signal events. 
The non-\CP mode sample, with relatively high statistics, helps constrain the PDF shapes for the low statistics \CP mode distributions.
The \DeltaE sideband sample helps define the $\mathcal{A}$ background~shape.

Since the values of $\xi$ obtained for each data sample were found to be 
consistent with each other, albeit with large statistical uncertainties,
we have constrained $\xi$ to have the same value for all data samples in the fit.
The simulation shows that the use of the same Gaussian parameters for all signal modes introduces only negligible systematic corrections. 
We assume that the \B decays found in the $m_{\Dz}$ sideband have the same final states as the signal and
we fit the same Gaussian to the doubly-peaking \B background.\par

\begin{figure}[t!]
\begin{center}
\setlength{\unitlength}{1mm}
\begin{picture}(80,150)(0,0)
\jput(0,0){\epsfig{file=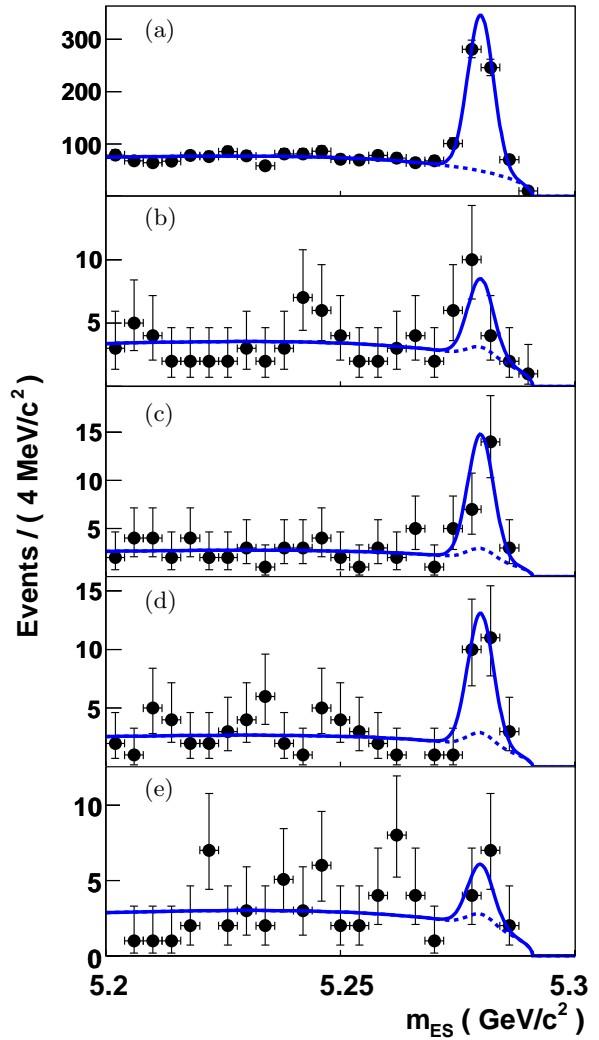,width=10cm}}
\jput(20,140){(a)}
\jput(20,115){(b)}
\jput(20,89){(c)}
\jput(20,64){(d)}
\jput(20,39){(e)}
\end{picture}

\caption{
Distributions of \mes in the signal region for
the non-\CP modes in \Bpm decays (a),
the \cpp modes in \Bp (b) and \Bm (c) decays and  
the \cpm modes in \Bp (d) and \Bm (e) decays. 
The dashed curve indicates the contribution from
the doubly-peaking \B-background estimated from 
a simultaneous fit to the \Dz sideband (not shown).
}
\label{fig:asymmetryfit}
\end{center}
\end{figure}

The doubly-peaking \B-background is assumed to not violate \CP and is therefore split equally between the \Bm and \Bp\ sub-samples. 
This assumption is considered further when we discuss the systematic uncertainties.
The fit results are shown graphically in Fig.~\ref{fig:asymmetryfit} and numerically
in Table~\ref{tab:nominalfitresult}.\par

\begin{table}[htbp]
\begin{center}
\caption{
\label{tab:nominalfitresult}
Results from the fits. For each \Dz mode class, we give the event
yield, the peaking background contribution $N_{{\rm peak}}$, \Acp and \Rcp . 
The uncertainties are statistical only.}
\begin{tabular}{ l c c c c}
\hline \hline
            & Yield                &$N_{{\rm peak}}$       &\Acp              &\Rcp              \\
\hline 
non-\CP     &$489 \pm 27$          &$ 20.1 \pm 5.6$  &                  &                  \\
\cpp\       &$\ 37.6 \pm\ 7.4$     &$\ 4.1 \pm 1.3$  &$-0.08 \pm 0.19$  &$\  1.96 \pm 0.40$\\
\cpm\       &$\ 14.8 \pm\ 5.9$     &$\ 3.0 \pm 2.3$  &$-0.24 \pm 0.35$  &$\ 0.73 \pm 0.29$ \\
\hline \hline
\end{tabular}
\end{center}
\end{table}
The statistical significance of the \cpp and \cpm yields are 6.8 and 2.7 standard deviations, respectively.
The yields for each individual mode are $23.1\pm5.1$ ($\Kp\Km$), $17.4\pm5.0$ ($\pip\pim$), 
$10.9\pm4.1$ ($\KS\piz$), $3.1\pm3.2$ ($\KS\phi$) and $3.8\pm2.7$ ($\KS\omega$).\par
Although most systematic errors cancel for \Acp,
an asymmetry inherent to the detector or data processing may exist. After performing the analysis on a high statistics
$\Bm\to\Dz\pim$ sample (not applying the \Kstarm selection), the final sample shows an asymmetry of $-0.019\pm0.008$.
We assign a systematic uncertainty of $\pm 0.027$. 
The second substantial systematic effect is a possible \CP~asymmetry in the peaking background.
Although there is no physics reason that requires the peaking background to be asymmetric, it cannot be excluded. 
We note that if there were an asymmetry $\mathcal{A}_{{\rm peak}}$, a systematic error on \Acp\ would be given
by $\mathcal{A}_{{\rm peak}} \times \frac{b}{c}$, where $b$ is the contribution of the peaking background and $c$ the signal yield. 
Assuming conservatively $|\mathcal{A}_{{\rm peak}}| \leq 0.5$, we obtain systematic errors 
of $\pm 0.06$ and $\pm 0.10$ on \Acpp\ and \Acpm respectively.\par 

Since \Rcp\ is a ratio of rates of processes with different final states of the \Dz, 
we must consider the uncertainties affecting the selection algorithms for the different $D$ channels. 
This results in small correction factors which account for the difference
between the actual detector response and the simulation model. The main effects
stem from the approximate modeling of the tracking efficiency (1.2\%
per track), the \KS\ reconstruction efficiency  for \cpm modes of
the \Dz\ (2.0\% per \KS), the \piz reconstruction efficiency for
the \KS\piz channel (3\%) and the efficiency and misidentification
probabilities from the particle identification (2\% per track). 
A substantial effect is the uncertainty on the measured branching fractions~\cite{bib:pdg2004}. 
Altogether, we obtain systematic uncertainties equal to $\pm 0.11$ and $\pm 0.055$ for
\Rcpp\ and \Rcpm~respectively.

Another systematic correction is applied to the \cpm measurements which arises from a possible \cpp 
background for the $\KS\phi$ and $\KS\omega$ channels. In this case, the observed
quantities, ${\cal A}^{{\rm obs}}_{\cpm}$ and ${\cal R}^{{\rm obs}}_{\cpm}$ are corrected: 
\begin{eqnarray}
\Acpm = ( 1 + \epsilon ) {\cal A}^{{\rm obs}}_{\cpm} - \epsilon  \Acpp ; \label{eq:Acorr} \ 
\Rcpm =  \frac{{\cal R}^{{\rm obs}}_{\cpm} }{ ( 1 + \epsilon ) }, \label{eq:Rcorr} \nonumber
\end{eqnarray}

\noindent
where $\epsilon$ is the ratio of \cpp background to \cpm signal.
There is little information on this \cpp background. 
An investigation in \babar\ of the $\Dz\to\Km\Kp\KS$ Dalitz plot indicates that the 
dominant background for $\Dz\to\KS\phi$ comes from the decay $a_0(980)\to\Kp\Km$, at the level of 
$(25\pm 1)\%$ the size of the $\phi\KS$ signal.
We have no information for the $\omega\KS$ channel and assume $(30\pm30)\%$.
Adding the most frequent $\KS\piz$ mode which does not suffer such a \cpp pollution, we estimate  $\epsilon = (13\pm7)$\%. 
The systematic error associated with this correction is $\pm0.01$ and $\pm0.04$ for \Acpm and \Rcpm, respectively.\par
To account for the non-resonant $\KS \pim$ pairs under the \Kstarm, we vary by $2\pi$ all the strong phases in a conservative model which incorporates 
S-wave $K\pi$ pairs in both $b\to c\bar{u}s$ and $b\to u\bar{c}s$ amplitudes.
This background 
induces systematic 
variations of $\pm 0.051$ for ${\cal A}_{\CP \pm}$ and $\pm 0.035$ for ${\cal R}_{\CP \pm}$. 
We add the systematic uncertainties in quadrature and quote the final results:

\vspace*{-0.2in}
\begin{eqnarray}
{\cal A}_{\CP+} &=& -0.08 \pm 0.19\ {\rm (stat.)} \pm 0.08\ {\rm (syst.)}, \nonumber\\
{\cal A}_{\CP-} &=& -0.26 \pm 0.40\ {\rm (stat.)} \pm 0.12\ {\rm (syst.)}, \nonumber\\
{\cal R}_{\CP+} &=& ~~1.96\pm 0.40\ {\rm (stat.)} \pm 0.11\ {\rm (syst.)}, \nonumber\\
{\cal R}_{\CP-} &=& ~~0.65\pm 0.26\ {\rm (stat.)} \pm 0.08\ {\rm (syst.)}. \nonumber
\end{eqnarray}
\vspace*{-0.2in}

\noindent
These results can also be expressed in terms of $x^\pm$ defined in equation~\ref{xvariables}:
\begin{eqnarray}
 x^+ &=& 0.32 \pm 0.18\ {\rm (stat.)} \pm 0.07\ {\rm (syst.)}, \nonumber \\
 x^- &=& 0.33 \pm 0.16\ {\rm (stat.)} \pm 0.06\ {\rm (syst.)}, \nonumber
\end{eqnarray}

\noindent
where the \cpp pollution systematic effects increase $x^+$ and $x^-$ by $0.022 \pm 0.012$ and $0.019 \pm 0.010$ respectively.
From equation~(\ref{eq:cpinfo-rcp}) we find $r_{\B}^2=0.30 \pm 0.25$.
\par

In summary, we have studied the decays of charged \B-mesons to a
$\Kstar\mathrm{ (892)}^-$ and a \Dz, where the latter is seen in final states of even and odd \CP . We express the
results with \Rcp,  \Acp and $x^\pm$. These quantities can be combined
with other  $D^{(*)}K^{(*)}$ measurements to estimate $r_B$ more precisely and improve our understanding of the angle $\gamma$.

We are grateful for the excellent luminosity and machine conditions
provided by our \pep2\ colleagues, 
and for the substantial dedicated effort from
the computing organizations that support \babar.
The collaborating institutions wish to thank 
SLAC for its support and kind hospitality. 
This work is supported by
DOE
and NSF (USA),
NSERC (Canada),
IHEP (China),
CEA and
CNRS-IN2P3
(France),
BMBF and DFG
(Germany),
INFN (Italy),
FOM (The Netherlands),
NFR (Norway),
MIST (Russia), and
PPARC (United Kingdom). 
Individuals have received support from CONACyT (Mexico), A.~P.~Sloan Foundation, 
Research Corporation,
and Alexander von Humboldt Foundation.


\begin{thebibliography}{99}

\bibitem{bib:theory} 
M. Gronau and D. London,
\plb{253}, 483 (1991).
M.~Gronau and D.~Wyler,
\plb{265}, 172 (1991).
A. B.Carter and A. I. Sanda, \jprd{23},7, 1567 (1981).
I. I. Bigi and A. I. Sanda , \plb{211}, 213 (1988).
D.~Atwood, I.~Dunietz, A.~Soni, \jprl{78}, 3257 (1997).
A.~Giri, Y.~Grossman, A.~Soffer and J.~Zupan,
Phys.\ Rev.\ D {\bf 68}, 054018 (2003).
\bibitem{bib:gronau}
M.~Gronau, 
\jprd{58}, 037301 (1998).
\bibitem{Aubert:2003uy}
  \babar\ Collaboration, B.~Aubert {\it et al.},
  Phys.\ Rev.\ Lett.\  {\bf 92}, 202002 (2004).
\bibitem{Aubert:2004fb}
  \babar\ Collaboration, B.~Aubert {\it et al.},
  Phys.\ Rev.\ Lett.\  {\bf 93}, 131804 (2004)
\bibitem{Aubert:2004hu}
  \babar\ Collaboration, B.~Aubert {\it et al.},
  Phys.\ Rev.\ D {\bf 71}, 031102 (2005).
\bibitem{Swain:2003yu}
  Belle Collaboration, S.~K.~Swain {\it et al.},
  Phys.\ Rev.\ D {\bf 68}, 051101 (2003).
\bibitem{Poluektov:2004mf}
  Belle Collaboration, A.~Poluektov {\it et al.},
  Phys.\ Rev.\ D {\bf 70}, 072003 (2004).
\bibitem{Gronau:2002mu}
  M.~Gronau,
  Phys.\ Lett.\ B {\bf 557}, 198 (2003).
\bibitem{Charles:2004jd}
  CKMfitter Group, J.~Charles {\it et al.},
  Eur.\ Phys.\ J.\ C {\bf 41}, 1 (2005).
\bibitem{bib:babar}
\babar\ Collaboration, B.\ Aubert {\em et al.},
\nima{479}, 1 (2002).
\bibitem{bib:pdg2004} Particle Data Group,
S. Eidelman {\em et al.}, 
\plb{592}, 1 (2004).
\bibitem{bib:dalitzomega} 
The \babar\ Collaboration, B.\ Aubert {\em et al.},
\jprd{69}, 032004 (2004).
\bibitem{bib:fisher} R. A. Fisher, Annals Eugen. {\bf 7}, 179 (1936).
\bibitem{bib:muriel}
\babar\ Collaboration, B.\ Aubert {\em et al.}, \jprl{89}, 281802 (2002).
\bibitem{bib:argus} 
The function is $\mathcal{A} (\mes) \propto \mes \sqrt{1-x^2} exp[-\xi (1-x^2)]$, where $x=2 \mes / \sqrt{s}$
and $\xi$ is a fit parameter; ARGUS Collaboration, H. Albrecht {\em et al.},  
\plb{185}, 218, (1987); ibid. \ {\bf 241}, 278, (1990).
\end{thebibliography}
\end{document}